\documentclass[10pt]{article}
\usepackage{graphicx}

\def\Title#1{\begin{center} {\Large #1 } \end{center}}
\def\Author#1{\begin{center}{ \sc #1} \end{center}}
\def\Address#1{\begin{center}{ \it #1} \end{center}}

\newcommand\pubblock{\rightline{\begin{tabular}{l} Proceedings of the Second Annual LHCP\\ \pubnumber\\
         \pubdate  \end{tabular}}}

\newenvironment{Abstract}{\begin{quotation} \begin{center} 
             \large ABSTRACT \end{center}\bigskip 
      \begin{center}\begin{large}}{\end{large}\end{center} \end{quotation}}

\newenvironment{Presented}{\begin{quotation} \begin{center} 
             PRESENTED AT\end{center}\bigskip 
      \begin{center}\begin{large}}{\end{large}\end{center} \end{quotation}}

\def\beq{\begin{equation}}
\def\eeq#1{\label{#1}\end{equation}}
\def\eeqn{\end{equation}}

\def\beqa{\begin{eqnarray}}
\def\eeqa#1{\label{#1}\end{eqnarray}}
\def\eeqan{\end{eqnarray}}

\let\bar=\overbar

\def\Dslash{\not{\hbox{\kern-4pt $D$}}}
\def\dslash{\not{\hbox{\kern-2pt $\del$}}}

\def\msb{{\bar{\ssstyle M \kern -1pt S}}}

\usepackage{color}

\newcommand\pubnumber{ CMS CR-2014/199 }

\newcommand\pubdate{\today}

\def\affiliation{
On behalf of the CMS Experiment, \\
Centre for Cosmology, Particle Physics and Phenomenology - CP3,\\
Universit\'e catholique de Louvain, Louvain-la-Neuve, Belgium}

\begin{document}

\large
\begin{titlepage}
\pubblock

\vfill
\Title{ Future prospects of Higgs Physics at CMS }
\vfill

\Author{ Miguel Vidal }
\Address{\affiliation}
\vfill
\begin{Abstract}

The Higgs boson physics reach of the CMS detector with 300(0)~fb$^{-1}$ of
proton-proton collisions at $\sqrt{s} = 14$~TeV is presented.
Precision measurements of the Higgs boson properties, Higgs boson pair
production and self-coupling, rare Higgs boson decays, and the potential for
additional Higgs bosons are discussed.

\end{Abstract}
\vfill

\begin{Presented}
The Second Annual Conference\\
 on Large Hadron Collider Physics \\
Columbia University, New York, U.S.A \\ 
June 2-7, 2014
\end{Presented}
\vfill
\end{titlepage}
\def\thefootnote{\fnsymbol{footnote}}
\setcounter{footnote}{0}
%

\normalsize 


\section{Introduction}

The standard model (SM) predicts the existence of a Higgs boson
responsible for the spontaneous electroweak symmetry
breaking~\cite{Phiggs, Englert}. In July 2012, the ATLAS and CMS~\cite{cms} Collaborations announced the discovery of a
new boson~\cite{higgsATLAS, higgsCMS} with a mass around 126 GeV. From
that moment one of the highest priorities of the CMS Collaboration was to
establish the nature of this new partible by studying its
properties using the current dataset. In parallel, the physics
goals of the future LHC running are clear~\cite{CMS:2013xfa}, including the measurement
of the Higgs boson properties to the highest precision, the Higgs
self-coupling, and the search for additional Higgs bosons and exotic
decays.

The LHC is currently in the first long shutdown in order prepare
for running in 2015. A second long shutdown in 2018 will be used
to upgrade de detectors for running at double the design luminosity
and an average number of interactions per pp crossing  (pile-up) of
50. The next phase of planned LHC operation,
referred to as the High Luminosity LHC (HL-LHC), will begin with the
third long shutdown in the period 2022-2023, where machine and
detectors will be upgraded to allow for pp running at a luminosity of
$5 \times 10^{34}$~cm$^{-2}$~s$^{-1}$,
with the goal of accumulating 3000~fb$^{-1}$.

In order to deal with the increased luminosity that the LHC will
deliver in the next runs, several modifications in the CMS detector
are required. The improvements are referred to as to ``Phase 1'' and
``Phase 2'' upgrades.

The results presented in this document are extrapolated to 300 and 3000~fb$^{-1}$ at $\sqrt{s} =
14$~TeV by scaling signal and background event yields. In the
particular case of the precision measurements, the extrapolations were
made assuming that future CMS upgrades will provide the same level of
detector and trigger performances achieved with the current
detector. Two different scenarios are presented. In Scenario 1, all
systematic uncertainties are left unchanged. In Scenario 2, the
theoretical uncertainties are scaled by a factor 1/2, while other
systematic uncertainties are scaled by the squared root of the
integrated luminosity.

\section{SM Higgs precision measurements}

\begin{figure}[!htb]
\centering
\includegraphics[height=2in]{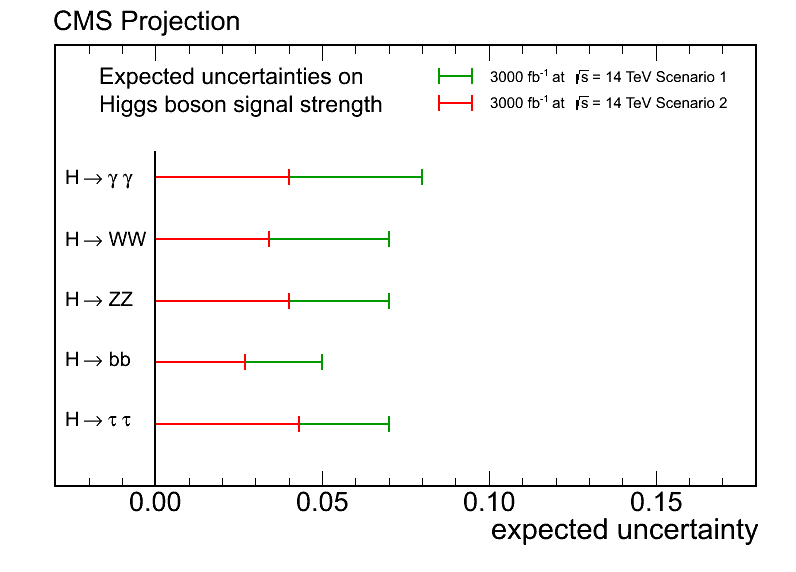}
\caption{Estimated precision on the measurements of the signal
  strength for a SM-like Higgs boson. The projections assume $\sqrt{s} =
14$~TeV and an integrated dataset of 3000~fb$^{-1}$.}
\label{fig:precision1}
\end{figure}

The signal strength modifier $\mu = \sigma/ \sigma_{SM}$, obtained in the combination of all search channels,
provides a first compatibility test. Figure~\ref{fig:precision1} shows
the $\mu$ uncertainties obtained in different sub-combinations of search channels, organized by decay modes for an integrated
dataset of 300~fb$^{-1}$ and 3000~fb$^{-1}$. We predict a precision
6–14\% for 300 fb$^{-1}$ and 4-8\% for a dataset of
3000~fb$^{-1}$. Studies show that future measurements of the signal
strength will be limited by theoretical uncertainty of the signal cross section, which is included in the fit.

The event yield for any (production) $\times$ (decay) mode is related to the production cross section
and the partial and total Higgs boson decay widths via the
narrow-width approximation:

\begin{equation}
 (\sigma \times BR) (x \to H \to {\it ff}) = \frac{\sigma_{x} \cdot \Gamma_{ff}}{\Gamma_{total}},
\end{equation}

where $\sigma_{x}$ is the production cross section through the initial
state $x$, $\Gamma_{ff}$ is the partial decay width into the final
state $ff$, and $\Gamma_{tot}$ is the total width of the Higgs boson. 
The possibility of Higgs boson decays to BSM particles, with a partial width
$\Gamma_{BSM}$, is accommodated by keeping $\Gamma_{tot}$  as a dependent parameter so that
$\Gamma_{tot} = \sum \Gamma_{ii} + \Gamma_{BSM}$, where the
$\Gamma_{ii}$ stand for the partial width of decay to all SM particles. The partial widths are proportional
to the square of the effective Higgs boson couplings to the corresponding particles. To test
for possible deviations in the data from the rates expected in the different channels for the SM
Higgs boson, factors $k_{i}$ corresponding to the coupling modifiers are introduced and fit to the
data~\cite{hxswg}.

\begin{figure}[!htb]
\centering
\includegraphics[height=2in]{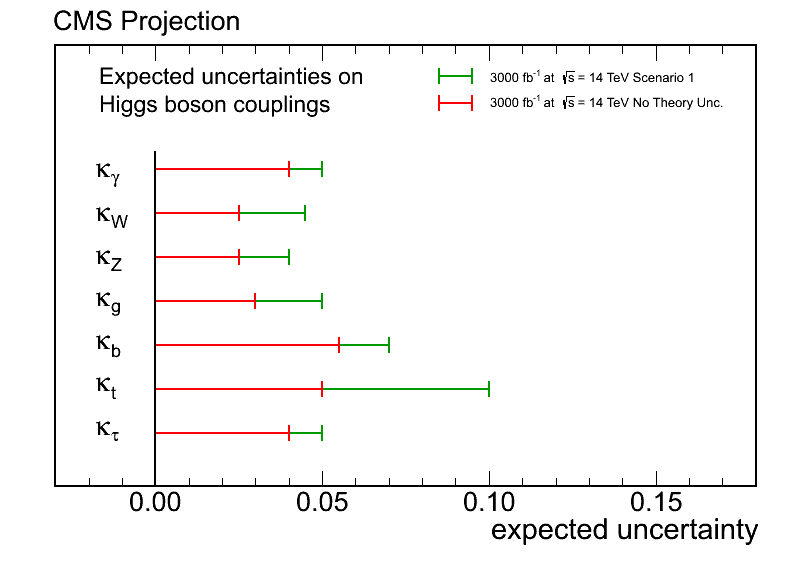}
\includegraphics[height=2in]{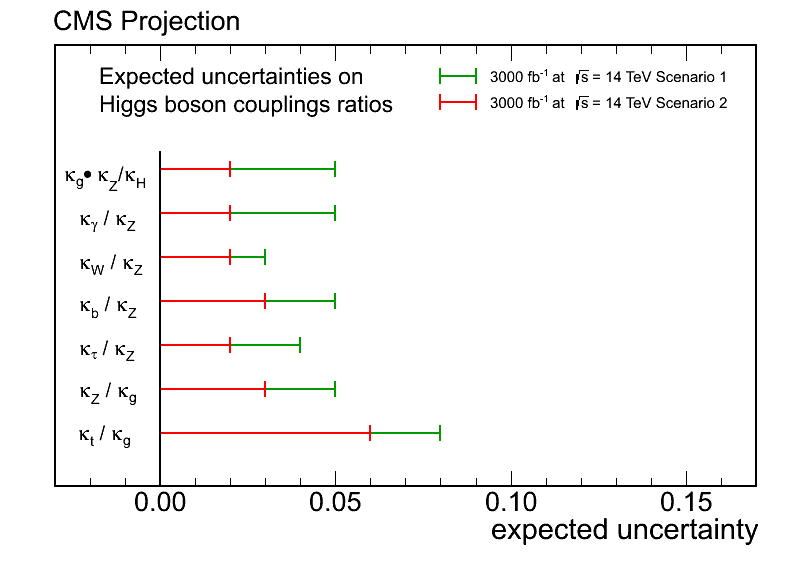}
\caption{Estimated precision on the measurements of $k_\gamma, k_W,
  k_Z, k_g, k_b, k_t$, and $k_\tau$ (right) and its ratios (left). The
projections assumes $\sqrt{s} = 14$~TeV and an integrated dataset of 3000~fb$^{-1}$.} 
\label{fig:precision2}
\end{figure}

Figure~\ref{fig:precision2}  shows the uncertainties obtained on
$k_{i}$ for an integrated dataset of 3000~fb$^{-1}$. The expected
precision ranges from 2-10\%. The measurements will be limited by systematic uncertainties on the cross section,
which is included in the fit for the signal strength. The statistical
uncertainties on $k_{i}$ are below one percent.

Besides testing Higgs boson couplings, it is important to determine the spin and quantum numbers
of the new particle as accurately as possible. The full case study has been presented by CMS
with the example of separation of the SM Higgs boson model and the
pseudoscalar (0$^{-}$) model.
Studies on the prospects of measuring CP-mixing of the Higgs boson are presented using the
$H \to ZZ^{*} \to 4l$ channel. The decay amplitude for a spin-zero boson defined as

\begin{equation}
  A(H \to ZZ) = v^{-1} (a_{1} m^{2}_{Z} \epsilon^{*}_{1} \epsilon^{*}_{2} + a_{2} f^{*(1)}_{\mu\nu}  f^{*(2), \mu\nu}+
  a_{3} f^{*(1)}_{\mu\nu}  \tilde{f}^{*(2), \mu\nu},
\end{equation}

where $f^{(i), \mu\nu}, (\tilde{f}^{(i), \mu\nu})$ is the (conjugate) field strength tensor of a
Z boson with polarization vector $\epsilon_{i}$ and $v$ the vacuum
expectation value of the Higgs field. The spin-zero models $0^+$ and
$0^-$ correspond to the terms with $a_1$ and $a_3$, respectively.

Four independent real numbers describe the process in Eq. (2), provided that the overall rate
is treated separately and one overall complex phase is not measurable. For a vector-boson
coupling, the four independent parameters can be represented by two fractions of the corre-
sponding cross-sections ($f_{a2}$ and $f_{a3}$) and two phases
($\phi_{a2}$ and $\phi_{a3}$). In particular, the fraction of CP-odd
contribution is defined under the assumption $a_2=0$ as

\begin{equation}
f_{a3} = \frac{|a_3|^{2} \sigma_3}{|a_1|^{2} \sigma_1 + |a_3|^{2} \sigma_3},
\end{equation}

where $\sigma_i$ is the effective cross section of the process corresponding to
$a_i =1$, $a_{j \neq i} =0$. Given the measured value of $f_{a3}$, the
coupling constants can be extracted in any parameterization. 

\begin{figure}[htb]
\centering
\includegraphics[height=2in]{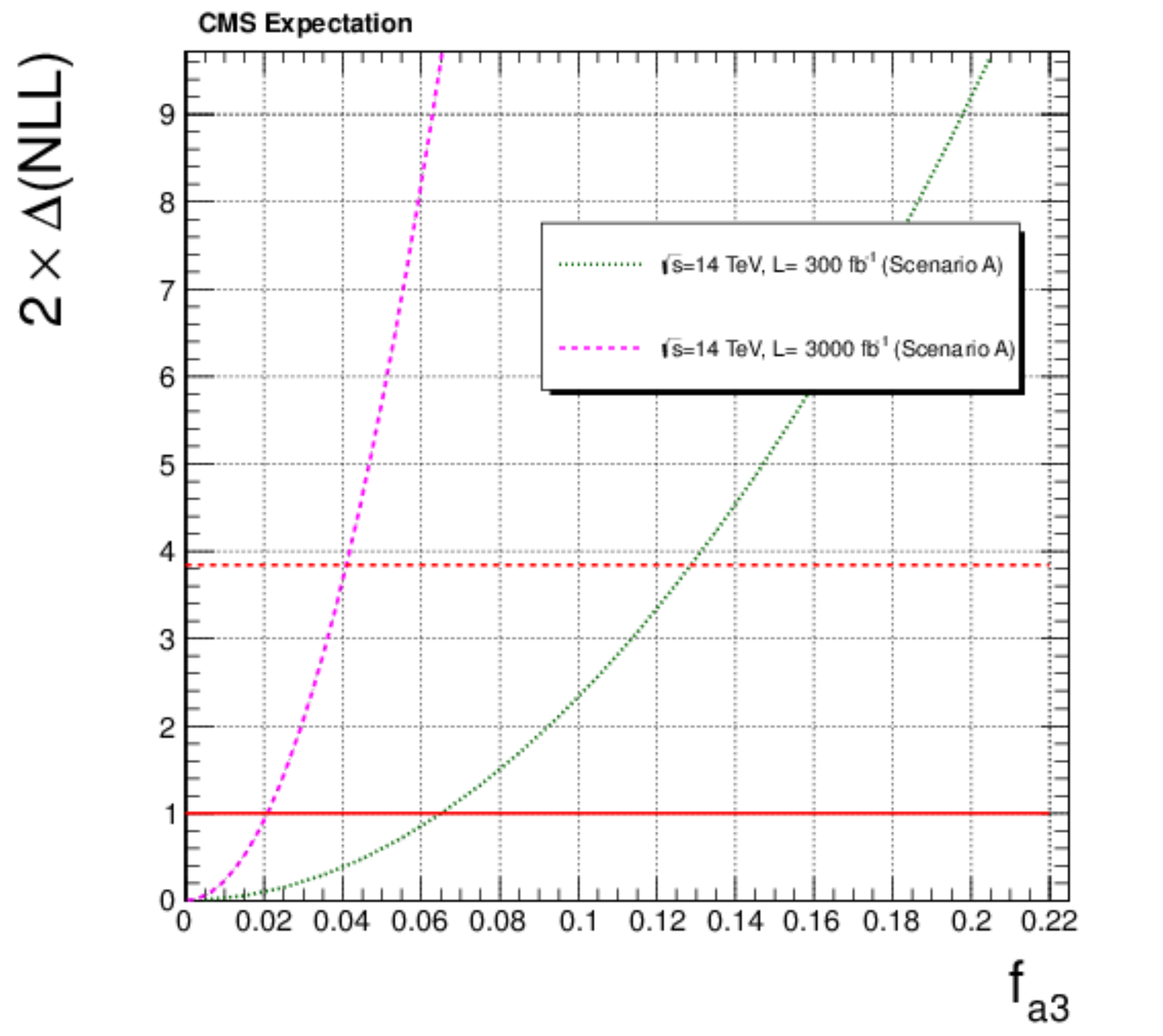}
\caption{ Distribution of expected $-2ln \mathcal{L}$ for $f_{a3}$ for
  the projection to 300~fb$^{-1}$ (green, dotted) and 3000~fb$^{-1}$ (magenta, dot-dashed). }
\label{fig:precision3}
\end{figure}

Projections of the expected $-2
ln \mathcal{L}$ values from the fits assuming 300~fb$^{-1}$ and
3000~fb$^{-1}$ are shown in figure~\ref{fig:precision3}.

\section{Higgs boson pair production and self-coupling}

In the SM, the Higgs boson mass itself fixes the value of the self-coupling 
in the scalar potential whose form is determined by the global symmetries and the requirement of renormalisability.
Direct information on the Higgs three- and four-point interactions could provide a indication of the 
scalar potential structure. To answer some of these questions the
Higgs boson pair production will play a major role. First through the simplest 
production process, sensitive to the self-coupling $\lambda$, and second probing the existence of heavier states 
coupled to the SM Higgs boson. However, even in most optimistic scenario in terms of energy and integrated
luminosity at the future HL-LHC, the measurement of the Higgs boson pair production remains extremely challenging.  
The SM cross section for HH production at 14~TeV is of the order of tens of 
femtobarns. Several new physics scenarios could enhance the HH production, opening the 
possibility for a BSM discovery before the SM HH measurement.

\section{SM rare decays}
We present two different cases of rare SM decays, $H \to \mu\mu $ and $H \to Z \gamma$. 

The dimuon decay channel has the advantage of a clean signature in the
CMS detector, but the disadvantage in the SM of a small branching ratio (BR), $2.2 \times 10^{-4}$ 
, for a SM Higgs boson with $m_H=125$GeV. The search is also
motivated by the fact that the BR of the SM Higgs boson decay in the
dimuon channel measures the second 
generation fermion Higgs-Yukawa coupling constant, the muon coupling,
$g_\mu$. Compared to the $H \to \gamma\gamma$ channel, the BR is ten
times smaller, but certain non-SM Higgs models predict enhanced branching ratios. 

\begin{figure}[htb]
\centering
\includegraphics[height=2in]{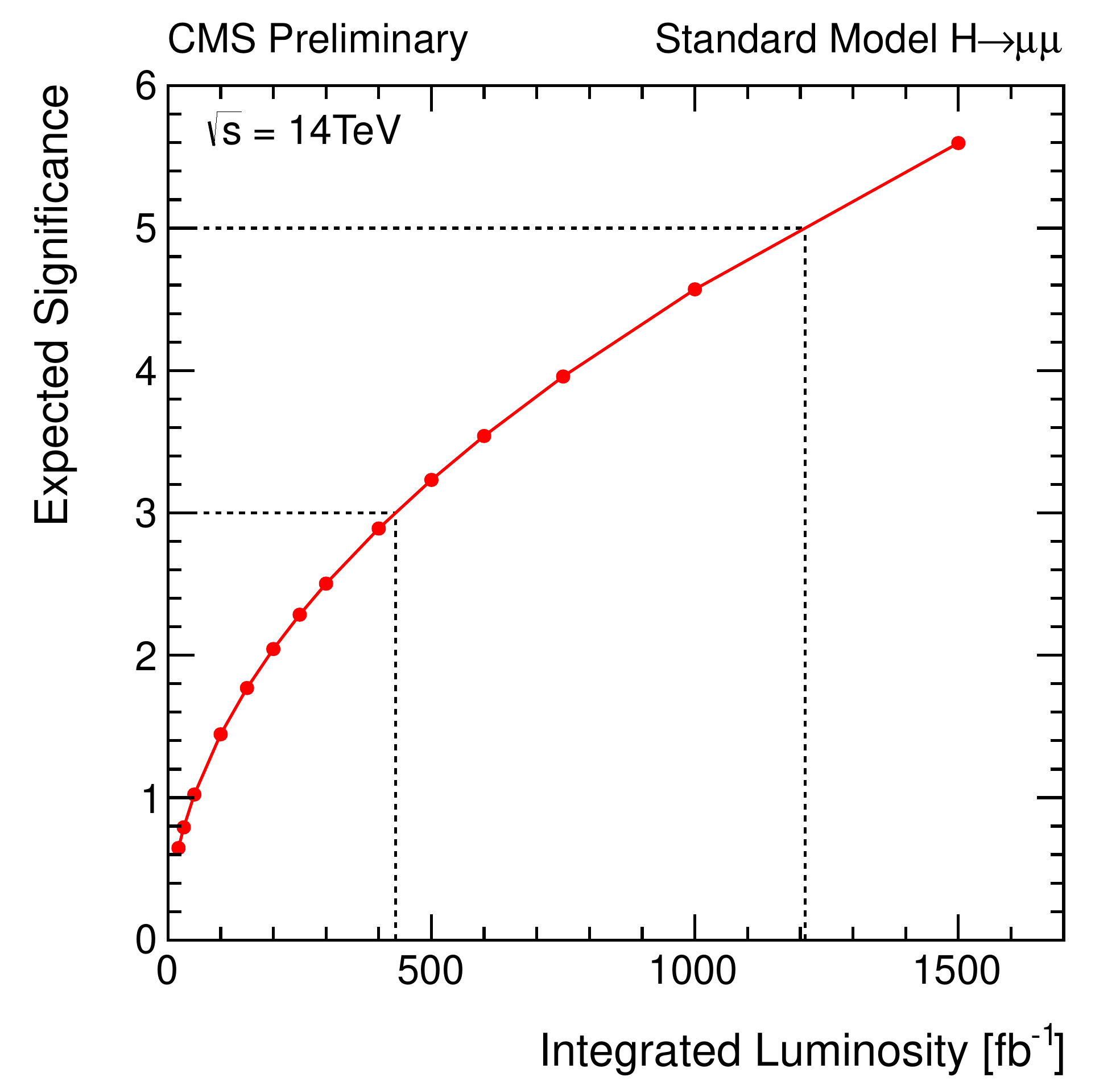}
\caption{ Expected significance versus luminosity for $\sqrt{s} = 14$~TeV }
\label{fig:raredecays}
\end{figure}

Figure~\ref{fig:raredecays} shows the 
extrapolated CMS $H\rightarrow\mu\mu$ search performance
for high luminosities at $\sqrt{s} = 14$~TeV from the current analyses
and $\sqrt{s} = 8$~TeV MC. No additional optimization for the higher
energy or luminosity is performed, and effects of higher pile-up,
detector upgrades, and detector aging are neglected. This is justified
by assuming that future degradation in detector performance will be 
counteracted by future analysis optimization. Based on this extrapolation a $5\sigma$ expected significance will be
reached with 1200~fb$^{-1}$.

The $H \to Z \gamma$ decay
channel is important because its partial width is induced by loops of heavy
charged particles, making it sensitive to physics beyond the standard model, just as the $H \to \gamma \gamma$ decay channel. Despite its small branching fraction, which in the SM varies between
0.111\% and 0.246\% in the mass range of $120 < m_H < 150$~GeV, the LHC experiments should be
sensitive to this channel in the near future.

\section{BSM rare decays}

The Higgs field is theorized to interact with all the fundamental particles, making
the particles massive (except the photon and gluon). One can postulate that there are additional
particles beyond the standard model that thus far escape our
detectors, to which the Higgs boson can
couple. This will enhance the Higgs boson invisible branching fraction significantly, compared to
$O$(0.1\%) as predicted in the SM via $H \to ZZ^* \to
4\nu$ decay. At a mass of 125~GeV, the invisible branching fraction of
the Higgs boson is especially
sensitive to new particles at electroweak scale. Many extensions to the SM, such as
supersymmetry, extra dimension models or dark matter singlet models, postulate
the existence of such invisible particles.

\section{Additional Higgs bosons}

Two Higgs doublet models (2HDM) provide an effective theory description for many
extensions of the electroweak symmetry breaking sector, allowing compact relations between
couplings of the observed Higgs boson and the production rates and branching fractions (BR)
of additional scalars. 2HDMs contain five physical Higgs bosons: two CP-even scalars h and
H, a CP-odd pseudo-scalar A, and a charged pair $H^{\pm}$. While the general parameter space of
2HDMs is large, it may be constrained by a variety of well-motivated assumptions about CP
conservation and the absence of new tree-level sources of flavor violation. Subject to these
constraints, 2HDMs may be parametrized in terms of nine variables: the physical masses
$m_h, m_H, m_A$, and $m_{H^{\pm}}$; the CP-even mixing angle $\alpha$;
the ratio of Higgs vacuum expectation values tan($\beta$); and three scalar couplings
$\lambda_5, \lambda_6$, and $\lambda_7$. The number of variables may
be further reduced by assuming the tree-level MSSM values of the three scalar couplings,
$\lambda_{5,6,7}$. The absence of tree-level flavor violation is guaranteed by four discrete choices of couplings
between SM fermions and the Higgs doublets; this leads to four types
of 2HDMs corresponding to the possible discrete coupling assignments. 
Of these, type I and II 2HDMs are the most familiar; the former can
give rise to a fermiophobic Higgs boson, while the latter includes
minimal supersymmetric extensions of the SM Higgs sector.

\begin{figure}[htb]
\centering
\includegraphics[height=2.5in]{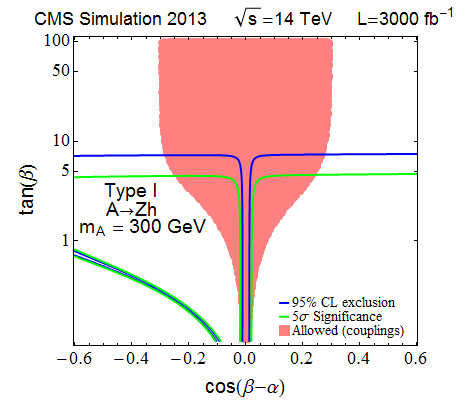}
\includegraphics[height=2.5in]{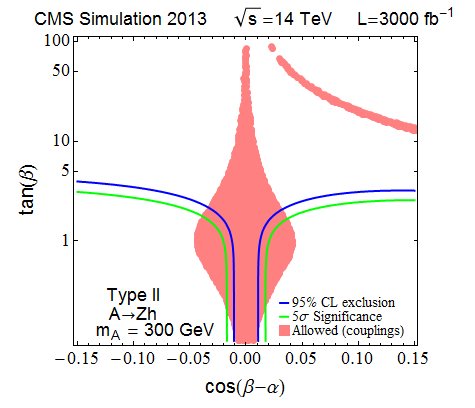}
\caption{ The region of parameter space for which a 300 GeV $A$ boson
  could be excluded at 95\% CL (blue), and the region of parameter space which could yield
  a 5$\sigma$ observation of a heavy pseudo-scalar $A$ boson (green) in the Zh channel, in the context
  of Type I (left) and II (right) 2HDMs. The colored regions correspond to the expected 95\% CL
  allowed region from Higgs boson precision measurements with 3000 fb$^{-1}$}
\label{fig:otherhiggs2}
\end{figure}

Here we extend a recent analysis completed in the context of the
Snowmass process, focusing on the resonant production and decay 
of the heavy CP-even scalar H boson and the pseudo-scalar $A$ boson, 
which possess large gluon fusion production cross sections at the LHC
as well as decays to distinctive final states such as $H \to ZZ$ and 
$A \to Zh$.   For these projections (figure~\ref{fig:otherhiggs2}) , we assume a total integrated
luminosity of 3000~fb$^{-1}$, collected at a center of mass
energy $\sqrt{s}$~TeV with an average of 140 pile-up interactions 
per bunch crossing. The expected performance of the so-called 
``configuration 3'' Compact Muon Solenoid (CMS) PhaseII 
detector upgrade proposal is used throughout. In configuration 3, 
the main upgrades consist of a new central silicon tracker, new
forward electromagnetic and hadronic calorimeters and a completed 
muon system, with a detection acceptance similar to that of the present CMS
detector.

\section{Conclusions}

We present results extrapolated to 300 and 3000~fb$^{-1}$ at $\sqrt{s} =14$~TeV. 
Precision measurements of the Higgs boson properties, Higgs boson pair
production and self-coupling, rare Higgs boson decays, and the potential for
additional Higgs bosons are discussed in the context of ``Phase 1'' and
``Phase 2''  CMS upgrades, aimed for the next LHC runs.


\end{document}